\RequirePackage{ifpdf}
\ifpdf 
\documentclass[pdftex]{sigma}
\else
\documentclass{sigma}
\fi

\numberwithin{equation}{section}
\numberwithin{figure}{subsection}
\numberwithin{example}{subsection}
\numberwithin{remark}{section}

\begin{document}
\allowdisplaybreaks
\renewcommand{\PaperNumber}{088}

\FirstPageHeading

\renewcommand{\thefootnote}{$\star$}

\ShortArticleName{Solvable Nonlinear Evolution PDEs in Multidimensional Space}

\ArticleName{Solvable Nonlinear Evolution PDEs\\ in Multidimensional Space\footnote{This paper is a contribution
to the Vadim Kuznetsov Memorial Issue ``Integrable Systems and Related Topics''.
The full collection is available at
\href{http://www.emis.de/journals/SIGMA/kuznetsov.html}{http://www.emis.de/journals/SIGMA/kuznetsov.html}}}

\Author{Francesco CALOGERO~$^\dag$ and Matteo SOMMACAL~$^{\ddag\S}$} %
\AuthorNameForHeading{F. Calogero and M. Sommacal}

\Address{$^\dag$~Dipartimento di Fisica, Universit\`{a} di Roma ``La
Sapienza'', \\
$\hphantom{^\dag}$ Istituto Nazionale di Fisica Nucleare, Sezione di Roma, \\
$\hphantom{^\dag}$~P.le Aldo Moro 2, 00185 Rome, Italy}
\EmailD{\href{mailto:francesco.calogero@roma1.infn.it}{francesco.calogero@roma1.infn.it},
\href{mailto:francesco.calogero@uniroma1.it}{francesco.calogero@uniroma1.it}}

\Address{$^{\ddag}$~Laboratoire J.-L. Lions, Universit\'{e} Pierre
et Marie Curie, Paris VI,\\
$\hphantom{^\ddag}$~175 Rue du Chevaleret, 75013
Paris, France {\rm (until October 30th, 2006)}}

\Address{$^{\S}$~Dipartimento di Matematica, Universit\`{a} di Perugia,\\
$\hphantom{^\S}$~Via Vanvitelli 1, 06123 Perugia, Italy {\rm (from November 1st, 2006)}}
\EmailD{\href{mailto:matteo.sommacal@pg.infn.it}{matteo.sommacal@pg.infn.it}}

\ArticleDates{Received October 31, 2006; Published online December 08, 2006}

\Abstract{A class of \textit{solvable} (systems of) nonlinear
evolution PDEs in \textit{multidimensional} space is discussed. We
focus on a rotation-invariant system of PDEs of Schr\"{o}dinger
type
and on a relativistically-invariant system of PDEs of Klein--Gordon type.
\textit{Isochronous} variants of these evolution PDEs are also
considered.}

\Keywords{nonlinear evolution PDEs in multidimensions; solvable
PDEs; NLS-like equations; nonlinear Klein--Gordon-like equations;
isochronicity}

\Classification{35G25; 35Q40; 37M05}

\begin{quote}
\it This article is dedicated to the memory of Vadim Kuznetsov,
with whom we spent several happy days during a Gathering of
Scientists held at the Centro Internacional de Ciencias in
Cuernavaca, and as well when he visited us in Rome, and when we met
at several other meetings.
\end{quote}


\section{Introduction}

Over a decade ago a class of \textit{C-integrable}~-- i.e.,
\textit{solvable} via a \textit{Change of variables} -- systems of PDEs
in \textit{multidimensional} space were identif\/ied \cite{C1994}. (A problem involving
nonlinear PDEs is considered \textit{solvable} if its solution can be
obtained by performing algebraic operations -- such as f\/inding the zeros of
a given polynomial -- and by solving \textit{linear} PDEs; of course only
seldom these operations can be performed \textit{explicitly}.) In the
present paper -- motivated by the scarcity of \textit{solvable} models of
nonlinear evolution PDEs in \textit{multidimensions} hence by the interest
of \textit{any} such model -- we study (a subclass of) these \textit{solvable}
 PDEs in more detail than it was done hitherto. We focus mainly on the
system of PDEs of Schr\"{o}dinger type
\begin{equation}
i \psi _{n,t}-\Delta  \psi _{n}+W( \vec{r})  \psi
_{n}=2 \sum_{ m=1,\, m\neq n}^{N}\frac{a+b \psi _{n} \psi _{m}-\big( \vec{
\nabla} \psi _{n}\big) \cdot \big( \vec{\nabla} \psi _{m}\big) }{\psi
_{n}-\psi _{m}},  \label{Schr}
\end{equation}
which is \textit{rotation-invariant} if $W(
\vec{r}) =W( r) $, and on the
\textit{relativistically-invariant} system of PDEs of
Klein--Gordon type%
\begin{equation}
\psi _{n,tt}-\Delta  \psi _{n}+M^{ 2} \psi _{n}=2 \sum_{m=1, \, m\neq
n}^{N}\frac{a+b \psi _{n} \psi _{m}+\psi _{n,t} \psi _{m,t}-\big( \vec{%
\nabla} \psi _{n}\big) \cdot \big( \vec{\nabla} \psi _{m}\big) }{\psi
_{n}-\psi _{m}}.  \label{KG}
\end{equation}
 \textit{Notation.} Here and throughout $N$ is an arbitrary
positive integer ($N\geq 2$); the index $n,$ as well as other
analogous indices (see below), range generally from $1$ to $N$; the
dependent variables $\psi _{n}\equiv \psi _{n}\left(
\vec{r}, t\right) $ are generally considered \textit{complex}
(although this is only mandatory for the f\/irst,~(\ref{Schr}), of
these two systems of PDEs); the space variable $\vec{r}$ is a
(\textit{real}) $S$-vector (with $S$ \textit{arbitrary}: for $S=3$,
$\vec{r}\equiv \left( x,y,z\right) $), and $r$ indicates its
modulus, $r^{ 2}=\vec{r}\cdot \vec{r} $; a dot sandwiched among two
$S$-vectors denotes the standard (\textit{rotation-invariant})
scalar product (for instance, for $S=3$, $\vec{r}\cdot
\vec{r}=x^{ 2}+y^{ 2}+z^{ 2}$); the ``potential'' $W\left(
\vec{r}\right) $ is an arbitrary function of the spatial
coordinate~$\vec{r}$ (we will often assume that it only depends on the modulus
$r$ of the $S$-vector $\vec{r}$); the constants~$a$ and $b,$ as well
as the ``mass'' parameter $M,$ are arbitrary (they might also
vanish); $t$ is the (\textit{real}) time variable; subscripted
independent variables always denote partial dif\/ferentiations;
$\vec{\nabla}$ respectively $\Delta \equiv \vec{\nabla} \cdot
\vec{\nabla}$ are the gradient respectively the Laplace operator in
$S$-dimensional space. \textit{Isochronous} variants of these
evolution PDEs are also considered.

In the following Section 2 we review tersely the general class of PDEs
treatable in this manner and the technique to solve them. In Sections 3
respectively 4 we treat in some detail the PDEs~(\ref{Schr}) respectively~(\ref{KG}),
describing various properties of their solutions and reporting
some representative examples, and we exhibit their \textit{isochronous}
versions. In Section~5 we take advantage of the electronic format of this
article to present a few animations displaying visually the time evolution
of a few solutions of some of these solvable (systems of) nonlinear
evolution PDEs: in this article we restrict these presentations to very few
cases, all with space dimensiona\-li\-ty less than three (we are of course aware
that the three-dimensional case is probably the most interesting one --
since we seem to live in a three-dimensional world -- but the presentation
of animations in a three-dimensional context is somewhat more tricky and we
therefore postpone the display of such examples to future articles we hope
to issue soon). In Section~6 we outline future directions of research.

\section{A class of solvable (systems of) PDEs\\ in multidimensional space}

In this section we review tersely the basic idea allowing to
identify a class of \textit{solvable} nonlinear evolution PDEs. The
interested reader will f\/ind a more detailed treatment in the paper
where this approach was introduced \cite{C1978}, and especially in
\cite{C2001} where this method is treated in considerable detail:
see~Section 2.3 of this book, and other references quoted
there in Section~2.N. These treatments focussed however on
ODEs rather than PDEs: the extension to PDEs is rather
straightforward, although it took some time to realize its
feasibility \cite{C1994} (see also \mbox{Exercise 2.3.4.2-5} in~\cite{C2001}).

Let $\Psi\left(\psi;\vec{r},t\right)$ be a (time- and
space-dependent) \textit{monic} polynomial of degree $N$ in the
variable~$\psi$, and denote by $\psi_{n}\left(\vec{r},t\right)$ its
$N$ zeros and by $\varphi_{m}\left(\vec{r},t\right)$ its $N$
coef\/f\/icients:
\begin{subequations}
\label{Ans}
\begin{gather}
\Psi \left( \psi ;\vec{r},t\right) =\prod\limits_{n=1}^{N}\left[
\psi -\psi _{n}\left( \vec{r},t\right) \right] ,  \label{Ansa}
\\
\Psi \left( \psi ;\vec{r},t\right) =\psi
^{ N}+\sum_{m=1}^{N}\varphi _{m}\left( \vec{r},t\right)  \psi
^{ N-m} .  \label{Ansb}
\end{gather}
\end{subequations}
Assume then that the time evolution of the dependent
variable $\Psi$ -- hence its dependence on the time and space
independent variables $t$ and $\vec{r}$, as well as its dependence
on the independent variable $\psi$ -- is characterized by a
\textit{linear} evolution PDE, which must of course be consistent
with the fact that $\Psi$ is, for \textit{all} time, a
\textit{monic} polynomial of degree $N$ in $\psi$, see (\ref{Ans}).
The \textit{linear} character of this PDE entails that the
time-evolution of the $N$ coef\/f\/icients
$\varphi_{m}\left(\vec{r},t\right)$, see (\ref{Ansb}), is as well
\textit{linear}, possibly \textit{explicitly} solvable (see below).
On the other hand the corresponding time evolution of the $N$ zeros
$\psi_{n}\left(\vec{r},t\right)$ will be \textit{nonlinear}, due to
the \textit{nonlinear} character of the mapping relating the $N$
zeros $\psi _{n}\left( \vec{r},t\right) $ to the $N$ coef\/f\/icients
$\varphi _{m}\left( \vec{r} ,t\right) $ of the polynomial $\Psi
\left( \psi ;\vec{r},t\right) $, see (\ref{Ans}). These (systems
of) \textit{nonlinear} evolution PDEs satisf\/ied by the dependent
variables $\psi _{n}\left( \vec{r},t\right) $ are those referred to
in the title of this paper: they are indeed generally
\textit{solvable} by taking advantage of the mapping, see
(\ref{Ans}), relating the $ N$ dependent variables $\psi _{n}\left(
\vec{r},t\right) $ to the $N$ functions $\varphi _{m}\left(
\vec{r},t\right) $.

In particular it is easily seen (using, if need be, the formulas
provided in Section 2.3 of \cite{C2001}), that to the
evolution PDE
\begin{equation}
i \Psi _{t}-\Delta  \Psi -V(\vec{r}) \left( \psi  \Psi _{\psi }-N \Psi
\right) +a \Psi _{\psi \psi }+b \left[ \psi ^{ 2} \Psi _{\psi \psi
}-N \left( N-1\right)  \Psi \right] =0  \label{EqPSISchr}
\end{equation}
satisf\/ied by $\Psi \left( \psi ;\vec{r},t\right) ,$ there
corresponds for the $N$ zeros $\psi _{n}\left( \vec{r},t\right) ,$
see (\ref{Ansa}), just the system of nonlinear evolution PDEs of
Schr\"{o}dinger type (\ref{Schr}) with
\begin{equation}
W\left( \vec{r}\right) =V(\vec{r})-2 \left( N-1\right)  b,  \label{W}
\end{equation}
while the corresponding evolution of the $N$ coef\/f\/icients $\varphi
_{m}\left( \vec{r},t\right) $, see (\ref{Ansb}), is clearly given by the
system of \textit{linear }evolution PDEs
\begin{gather}
 i \varphi _{m,t}-\Delta  \varphi _{m}+\left[ W\left(
\vec{r}\right)
-b \left( m+3\right) \right]  m \varphi _{m} \nonumber \\
\qquad{} = -a \left( N-m+2\right)  \left( N-m+1\right)  \varphi
_{m-2},\qquad m=1,\dots,N   \label{EqPhiSchr}
\end{gather}
with $\varphi _{-1}=0$ and $\varphi _{0}=1$ (see (\ref{Ansb})). Note
that this system of \textit{linear} PDEs is \textit{decoupled} if
the constant $a$ vanishes; as we shall see in the following section,
it can also be replaced by a~\textit{decoupled} system if the
constant $a$ does \textit{not} vanish, $a\neq 0,$ but the potential
$W\left( \vec{r}\right) $ is constant, $W(\vec{r})=C$, see Section
3.

Likewise to the evolution PDE
\begin{equation}
\Psi _{tt}-\Delta  \Psi -\mu ^{ 2} \left[ \psi  \Psi _{\psi
}-N \Psi \right] +a \Psi _{\psi \psi }+b \left[ \psi ^{ 2} \Psi
_{\psi \psi }-N \left( N-1\right)  \Psi \right] =0,
\label{EqPSIKG}
\end{equation}
 there corresponds for the $N$ zeros $\psi _{n}\left(
\vec{r},t\right) $ just the system of \textit{nonlinear} evolution
PDEs of Klein--Gordon type (\ref {KG}) with
\begin{equation*}
M^{ 2}=\mu ^{ 2}-2 \left( N-1\right)  b,  
\end{equation*}
while the corresponding evolution of the $N$ coef\/f\/icients
$\varphi_{m}\left(\vec{r},t\right)$ is clearly given by the system
of \textit{linear} evolution PDEs
\begin{gather}
 \varphi _{m,tt}-\Delta  \varphi _{m}+\left[ M^{ 2}-b \left(
m+3\right)
\right]  m \varphi _{m} \nonumber \\
\qquad = -a \left( N-m+2\right)  \left( N-m+1\right)  \varphi
_{m-2},\qquad m=1,\ldots,N, \label{EqPhiKG}
\end{gather}
 again with $\varphi _{-1}=0$ and $\varphi _{0}=1$ (see
(\ref{Ansb})): this system is \textit{decoupled} if the constant $a$
vanishes, and can be replaced by a \textit{decoupled} system even if
$a$ does \textit{not} vanish, see Section~4.

Having conveyed, tersely but hopefully clearly, the main idea of this
approach to identify \textit{solvable} systems of \textit{nonlinear}
evolution PDEs, we turn, in the next two sections, to the study of the two
systems of \textit{nonlinear} evolution PDEs (\ref{Schr}) and (\ref{KG}).

\bigskip

\section{Solvable system of nonlinear PDEs of Schr\"{o}dinger type}

In this section we investigate the system of \textit{nonlinear}
evolution equations (\ref{Schr}), f\/irstly by analytic techniques and
subsequently by reporting some of its solutions in a representative
set of cases.

\begin{remark} \label{Remark 3.1}
The ``mean f\/ield'' $\bar{\psi}\left(\vec{r},t\right)$
\begin{equation}
\bar{\psi}\left( \vec{r},t\right) =\frac{1}{N} \sum_{n=1}^{N}\psi
_{n}\left( \vec{r},t\right)  \label{MeanField}
\end{equation}
 satisf\/ies the \textit{linear} Schr\"{o}dinger equation%
\begin{equation*}
i \bar{\psi}_{t}-\Delta  \bar{\psi}+W\left( \vec{r}\right)
 \bar{\psi} =0.
\end{equation*}
\begin{proof}
Sum the nonlinear evolution PDEs (\ref{Schr}) over $n$ from $%
1$ to $N$, use (\ref{MeanField}) in the left-hand side, and notice
that the double sum in the right-hand side vanishes due to the
antisymmetry of the summand under the exchange of the two dummy
indices $n$ and $m$.
\end{proof}
\end{remark}

\begin{remark} \label{Remark 3.2}
If \textit{all} the $N$ coef\/f\/icients $\varphi_{m}$ vanish,
$\varphi_{m}=0$, then $\Psi =\psi ^{ N}$ (see (\ref{Ansb})), hence
\textit{all} its $N$ zeros $\psi _{n}$ correspondingly vanish,
$\psi_{m}=0$ (and more generally: if the f\/irst $M$
coef\/f\/icients~$\varphi_{m}$ vanish, $\varphi_{m}=0$ for $m=1,\dots ,M$, then $M$ of
the $N$ zeros of $\Psi$ vanish). Hence to a set of \textit{localized
}solutions $\psi_{n}\left(\vec{r},t\right)$ of the system of
nonlinear evolution equations (\ref{Schr}), characterized by the
asymptotic conditions
\begin{subequations}
\label{Asymptpsiphi}
\begin{equation}
\underset{r\rightarrow \infty }{\lim }\left[ \psi _{n}\left(
\vec{r},t\right) \right] =0,\qquad n=1,\dots ,N,  \label{Asymtpsiphia}
\end{equation}
 there correspond a set of \textit{localized} solutions of
the system of \textit{linear} Schr\"{o}dinger PDEs (\ref{EqPhiSchr})
(with~(\ref{W})), characterized by the analogous asymptotic conditions%
\begin{equation}
\underset{r\rightarrow \infty }{\lim }\left[ \varphi _{n}\left( \vec{r}%
,t\right) \right] =0, \qquad n=1,\dots ,N;  \label{Asymtpsiphib}
\end{equation}
\end{subequations}
 and, of course, viceversa, namely clearly
(\ref{Asymtpsiphib}) entails (\ref{Asymtpsiphia}). However when two
\textit{different} components, $\psi_{n}\left(\vec{r}, t\right)$
and $\psi_{m}\left( \vec{r}, t\right)$ with $n\neq m$, of the
system of nonlinear evolution equations~(\ref{Schr}) are equal,
$\psi_{n}\left(\vec{r}, t\right)=\psi_{m}\left(\vec{r}, t\right)$,
clearly this system runs into a singularity due to the vanishing of
one of the denominators in the right-hand side of (\ref{Schr}); and
this situation gets even worst if the two \textit{different}
components, $\psi _{n}\left( \vec{r}, t\right) $ and
$\psi_{m}\left( \vec{r} , t\right) $ with $n\neq m$, both
\textit{vanish},
$\psi_{n}\left(\vec{r}, t\right)=\psi_{m}\left(\vec{r}, t\right)=0$.
Hence in all the examples considered below we shall try and avoid
this problem, and in particular we shall focus, rather than on
\textit{localized }solutions $\psi_{n}\left(\vec{r}, t\right)$ that
vanish asymptotically (see (\ref{Asymtpsiphia})), either on
solutions that oscillate asymptotically, or on solutions that are
asymptotically constant,
\begin{equation}
\underset{r\rightarrow \infty }{\lim }\left[ \psi _{n}\left( \vec{r}%
, t\right) \right] =a_{n}\qquad \text{with }a_{n}\neq a_{m}\quad \text{if} \quad n\neq
m.  \label{Asy}
\end{equation}
Note that these asymptotic values $a_{n}$ might depend on
the direction along which the space coordinate $\vec{r}$ diverges.
\end{remark}

\begin{remark} \label{Remark 3.3}
If the potential $V\left( r\right) $ is constant,
\begin{equation}
V\left( r\right) =B,  \label{Vconst}
\end{equation}
(entailing that $W\left(r\right)$ is as well
\textit{constant},
\begin{equation*}
W\left( r\right) =B-2 \left( N-1\right)  b,
\end{equation*}
see (\ref{W})), it may be convenient to replace the
expression (\ref{Ansb}) of the monic polynomial
$\Psi\left(\psi;\vec{r},t\right)$ in terms of its coef\/f\/icients
$\varphi_{m}\left(\vec{r},t\right)$ by the following representation:
\begin{subequations}
\label{PSIGegen}
\begin{gather}
\Psi \left( \psi ;\vec{r},t\right) =\frac{C_{N}^{ \gamma }\left( c \psi
\right) }{k_{N} c^{ N}}+\sum_{m=1}^{N}\chi _{m}\left( \vec{r},t\right)  %
\frac{C_{N-m}^{ \gamma }\left( c \psi \right) }{k_{N-m} c^{ N-m}},
\label{PSIGegena}
\\
c=\left( -\frac{b}{a}\right) ^{ 1/2},  \label{PSIGegenb}
\\
\gamma =-\frac{B+b}{2 b},  \label{PSIGegenc}
\end{gather}
\end{subequations}
where the polynomial $C_{\ell }^{ \gamma }\left( z\right)$, of
degree $ \ell $, is the standard Gegenbauer polynomial \cite{E2},
satisfying the ODE
\begin{equation*}
\left( 1-z^{ 2}\right)  C_{l}^{ \gamma  \prime \prime }\left(
z\right) -\left( 2 \gamma +1\right)  z C_{l}^{ \gamma  \prime
}\left( z\right) +\ell  \left( \ell +2 \gamma \right)
 C_{l}^{ \gamma }\left( z\right) =0 
\end{equation*}
(where appended primes denote derivatives with respect to
the argument of the function they are appended to, in this case with
respect to $z$), and being characterized by the asymptotic behavior
\begin{equation*}
\underset{z\rightarrow \infty }{\lim }\left[ \frac{C_{l}^{ \gamma }\left(
z\right) }{k_{\ell }}\right] =1, \qquad 
\mbox{where}\quad
k_{\ell }=\frac{2^{ \ell } \Gamma \left( \gamma +\ell \right) }{\ell
! \Gamma \left( \gamma \right) }.  
\end{equation*}
Then the requirement that
$\Psi\left(\psi;\vec{r},t\right)$ satisfy the PDE (\ref{EqPSISchr})
with (\ref{Vconst}), and therefore that its $N$ zeros
$\psi_{n}\left(\vec{r},t\right)$ satisfy the system of nonlinear
PDEs (\ref{Schr}) with (\ref{Vconst}), entails that the $N$
coef\/f\/icients $\chi_{m}\left(\vec{r},t\right)$ satisfy the (system of
\textit{decoupled}) \textit{linear} Schr\"{o}dinger\ PDEs (with
constant potentials)
\begin{equation}
i \chi _{m,t}-\Delta  \chi _{m}+\left[ B-b (2 N-m-1)\right]  m \chi
_{m}=0,\qquad m=1,\dots ,N.  \label{Decoupled}
\end{equation}
 The investigation of the limiting cases in which some
coef\/f\/icients vanish, and therefore a dif\/ferent set of classical
polynomials come into play in place of the Gegenbauer polynomials,
is left as an exercise for the diligent reader.
\end{remark}

The \textit{nonlinear} mapping among the $N$ dependent variables $\psi
_{n}\left( \vec{r},t\right) $ satisfying the system of \textit{nonlinear}
PDEs (\ref{Schr}) and the $N$ functions $\varphi _{m}\left( \vec{r},t\right)
$ respectively $\chi _{m}\left( \vec{r},t\right) $ satisfying the system of
\textit{linear} PDEs (\ref{EqPhiSchr}) respectively (\ref{Decoupled}),
entailed by the simultaneous validity of (\ref{Ansa}) and (\ref{Ansb})
respectively (\ref{PSIGegen}), is the key to the \textit{solvability} of the
system of \textit{nonlinear} PDEs (\ref{Schr}). This can be taken advantage
of in two ways: to solve the \textit{initial-value} problem for the system
of \textit{nonlinear} PDEs (\ref{Schr}), or to manufacture special, possibly
quite \textit{explicit}, solutions of this system of \textit{nonlinear} PDEs.

\bigskip

\subsection{Solution of the initial-value problem for the system\\ of
nonlinear Schr\"{o}dinger-like PDEs (\ref{Schr})}

The \textit{initial-value} problem consists in the determination of
the solution $\psi_{n}\left(\vec{r},t\right)$, $n=1,\dots ,N$, of the
system of Schr\"{o}dinger-like \textit{nonlinear} PDEs (\ref{Schr})
corresponding to \textit{assigned} initial data
$\psi_{m}\left(\vec{r},0\right)$, $m=1,\dots ,N$.

The \textit{first} step is to determine the corresponding initial
data $\varphi_{m}\left(\vec{r},0\right)$ of the system of
Schr\"{o}dinger-like \textit{linear} PDEs (\ref{EqPhiSchr}). This is
achieved by solving for the $N$ functions
$\varphi_{m}\left(\vec{r},0\right)$ the system
\begin{equation}
\psi ^{ N}+\sum_{m=1}^{N}\varphi _{m}\left( \vec{r},0\right)  \psi
^{ N-m}=\prod\limits_{n=1}^{N}\left[ \psi -\psi _{n}\left( \vec{r}
,0\right) \right] ,  \label{phinzero}
\end{equation}
entailed by the simultaneous validity of (\ref{Ansa}) and
(\ref{Ansb}) at $t=0$. This amounts to the determination of the $N$
coef\/f\/icients $\varphi_{m}$ of a monic polynomial given its $N$ zeros
$\psi_{n}$; the relevant, \textit{explicit} formulas are of course
well-known:
\begin{gather*}
\varphi_{1}\left(\vec{r},0\right)=
-\sum_{n=1}^{N}\psi_{n}\left(\vec{r},0\right) ,
\qquad
\varphi _{2}\left( \vec{r},0\right) =\sum_{n,m=1,\, m\neq n}^{N}\psi
_{n}\left( \vec{r},0\right)  \psi _{m}\left( \vec{r},0\right),
\end{gather*}%
and so on, up to
\begin{equation*}
\varphi _{N}\left( \vec{r},0\right) =\left( -\right)
^{ N} \prod\limits_{n=1}^{N}\psi _{n}\left( \vec{r},0\right).
\end{equation*}

The \textit{second} step is to solve the initial-value problem for
the system of evolution PDEs (\ref{EqPhiSchr}), obtaining thereby
its solution $\varphi_{m}\left(\vec{r},t\right)$ at time $t$. The
\textit{linear} character of this (coupled) system of evolution
PDEs, (\ref{EqPhiSchr}), provides the main simplif\/ication; of course
an \textit{explicit} solution is only possible for special choices
of the potential $W(\vec{r})$.

The \textit{third} step is to obtain the solution
$\psi_{n}\left(\vec{r},t\right)$ from the, now assumedly known,
functions $\varphi_{m}\left(\vec{r},t\right)$, via the relation
\begin{equation}
\prod\limits_{n}^{N}\left[ \psi -\psi _{n}\left( \vec{r},t\right)
\right] =\psi ^{ N}+\sum_{m=1}^{N}\varphi _{m}\left(
\vec{r},t\right)  \psi ^{ N-m},  \label{phipsi}
\end{equation}
again entailed by the simultaneous validity of (\ref{Ansa}) and
(\ref{Ansb}), but now at time $t$. This amounts of course just to
the purely algebraic task of f\/inding the zeros of a given monic
polynomial of degree $N$: an \textit{explicit} solution is of course
only possible for $N\leq 4$.

In the special case of a constant potential $V\left(r\right)$, see
(\ref{Vconst}), an alternative procedure of solution can be based on
the representation (\ref{PSIGegen}) rather than (\ref{Ansb}): this
eases the \textit{second} of the steps outlined above, but makes a
bit less simple the \textit{first} step. The diligent reader will
easily f\/igure out the relevant details.

\subsection{How to manufacture explicit solutions of the system \\ of nonlinear
Schr\"{o}\-din\-ger-like PDEs (\ref{Schr})}

Clearly the appropriate strategy -- underlining all the examples
exhibited below -- is to identify an \textit{explicit} solution
$\varphi_{m}\left(\vec{r},t\right)$ of the system of \textit{linear}
evolution PDEs (\ref{EqPhiSchr}), and then to obtain the
corresponding solution $\psi_{n}\left(\vec{r},t\right)$ of the
system of \textit{nonlinear} evolution PDEs (\ref{Schr}) via
(\ref{phipsi}), namely by identifying the $N$ zeros
$\psi_{n}\left(\vec{r},t\right)$ of the monic polynomial of degree
$N$ in $\psi$ having the coef\/f\/icients $\varphi
_{m}\left(\vec{r},t\right) $, see (\ref{Ansb}). This can of course
be done \textit{explicitly} only for $N\leq 4 $: not a~signif\/icant
restriction when it comes to the \textit{explicit} exhibition of
examples, which would indeed be impractical for larger values of $N$
(in Section~5 we indeed limit our exhibition of animations to the
$3$-body case, $N=3$).

An alternative route -- applicable when the potential is
\textit{constant}, see (\ref{Vconst}) -- takes as starting point an
\textit{explicit }solution $\chi_{m}\left(\vec{r},t\right)$ of the
(\textit{decoupled}) system of \textit{linear} evolution PDEs
(\ref{Decoupled}), and then obtains the corresponding solution
$\psi_{n}\left(\vec{r},t\right)$ of the system of \textit{nonlinear}
evolution PDEs~(\ref{Schr}) via (\ref{PSIGegen}), namely by
identifying the $N$ zeros $\psi_{n}\left(\vec{r},t\right)$ of the
monic polynomial of degree $N$ in $\psi$ given by this expression
(\ref{PSIGegen}).

\subsection{Examples}

\begin{example} \label{Example 3.3-1}
The simplest example is characterized by the assignment
\begin{equation*}
a=b=0,\qquad W(\vec{r})=V(\vec{r})=0,
\end{equation*}
 namely by the system of \textit{nonlinear} evolution PDEs
in one-dimensional space (see (\ref{Schr}))
\begin{equation}
i \psi _{n,t}-\Delta  \psi _{n}=-2 \sum_{m=1,\, m\neq n}^{N}\frac{\vec{
\nabla}\psi _{n}\cdot \vec{\nabla}\psi _{m}}{\psi _{n}-\psi _{m}}.
\label{SchrEx1}
\end{equation}

The corresponding system of \textit{linear} PDEs satisf\/ied by the
coef\/f\/icients $\varphi _{m}\left( \vec{r},t\right) $ of the
polynomial $\Psi \left( \psi ;\vec{r},t\right) $ in the variable
$\psi $ of which the solutions $\psi _{n}\left( \vec{r},t\right) $
are the $N$ zeros,
\begin{equation}
\Psi \left( \psi ;\vec{r},t\right) =\psi
^{ N}+\sum_{m=1}^{N}\varphi _{m}\left( \vec{r},t\right)  \psi
^{ N-m}=\prod\limits_{n}^{N}\left[ \psi -\psi _{n}\left(
\vec{r},t\right) \right] ,  \label{PSIPSI}
\end{equation}
 reads as follows (see (\ref{EqPhiSchr})):
\begin{equation}
i \varphi _{m,t}-\Delta \varphi _{m}=0,\qquad m=1,\dots ,N.  \label{Eqphi}
\end{equation}

A special class of ``traveling wave" solutions of these (decoupled)
\textit{linear} PDEs reads
\begin{equation}
\varphi _{m}\left( \vec{r},t\right) =\varphi _{m}\left(
\vec{r}-\vec{v}  t\right) =A_{m}+B_{m} \exp \left[
-i \vec{v} \left( \vec{r}-\vec{v}  t\right) \right] ,
\label{travelingwave}
\end{equation}
 with $\vec{v}$ an arbitrary \textit{real} constant
$S$-vector and the $2 N$ scalar constants $A_{m}$, $B_{m}$ also
arbitrary (possibly \textit{complex}). These solutions (which are
clearly the most general ones of ``traveling wave" character, namely
depending on the single $S$-vector $\vec{r}-\vec{v} t$ rather than
separately on the $S$-vector space variable $\vec{r}$ and
the scalar time variable $t$) are \textit{not} localized: they are
\textit{constant} along the $S-1$ space directions orthogonal to
$\vec{v}$, \textit{periodic} with period $L=\left\vert 2 \pi
 / v\right\vert $ along the space direction parallel to $\vec{v}$,
and \textit{periodic} in $t$ with period $T=2 \pi  v^{-2}$. The
corresponding \textit{traveling wave} solutions $\psi _{n}\left(
x,t\right) $ of the system of \textit{nonlinear} PDEs
(\ref{SchrEx1}) are the $N$ zeros of the following polynomial of
degree $N$ in the variable $\psi$:
\begin{subequations}
\label{Solpsi3.3-1}
\begin{equation}
P_{N}\left( \psi \right) +Q_{N-1}\left( \psi \right)  \exp \left[ -i \vec{v%
} \left( \vec{r}-\vec{v} t\right) \right]
=\prod\limits_{n}^{N}\left[ \psi -\psi _{n}\left( \vec{r},t\right)
\right],  \label{Solpsi3.3-1a}
\end{equation}
 where $P_{N}\left(\psi\right)$ is an arbitrary
\textit{monic} polynomial of degree $N$ and
$Q_{N-1}\left(\psi\right)$ is an arbitrary polynomial of degree
$N-1$,
\begin{equation}
P_{N}\left( \psi \right) =\psi ^{ N}+\sum_{m=1}^{N}A_{m} \psi
^{ N-m},\qquad Q_{N-1}\left( \psi \right) =\sum_{m=1}^{N}B_{m} \psi ^{ N-m}.
\label{P}
\end{equation}
\end{subequations}
 This entails of course that these solutions are as well
of the same traveling wave type, $\psi_{n}\left(\vec{r},t\right)
=\psi_{n}\left(\vec{r}-\vec{v} t\right)$, with the same periodicity
properties described above for $\varphi _{m}\left( \vec{r},t\right)
=\varphi_{m}\left( \vec{r}-\vec{v} t\right)$ -- or possibly with
time periods which are \textit{integer multiples} of that of the
coef\/f\/icients $\varphi_{m}\left(\vec{r},t\right)$ due to the
possibility that through the time evolution the zeros of the
polynomial (\ref{P}) exchange their roles; for each component
$\psi_{n}\left( \vec{r},t\right)$ this integer multiple cannot of
course exceed $N$, while for the entire solution
$\left\{\psi_{n}\left(\vec{r},t\right);n=1,\dots ,N\right\}$ the period
can be somewhat larger -- but generally not too much \cite{GS}.

A solution of this kind (with $S=2$ and $N=3,$ and a specif\/ic choice of the
remaining free parameters) is displayed as an animation in Section 5.

Another special set of solutions of (\ref{Eqphi}) -- written below,
for simplicity, for the one-dimensional case ($S=1$) -- reads
\begin{equation}
\varphi _{m}\left( x,t\right) =A_{m}+\frac{B_{m}}{\left( t+t_{m}+i \eta
_{m}\right) ^{ 1/2}} \exp \left[ \frac{-i \left( x-x_{m}+i \eta
_{m} v_{m}\right) ^{ 2}}{4 \left( t+t_{m}+i \eta _{m}\right) }\right] ,
\label{localizedphi}
\end{equation}
where the constants $A_{m}$ and $B_{m}$ are \textit{arbitrary}
(possibly \textit{complex}; but see below for some conditions on the
constants $A_{m}$), the constants $x_{m}$, $t_{m}$ and $v_{m}$ are
also arbitrary but \textit{real}, and the constants $\eta _{m}$ are
also arbitrary but \textit{positive}, $\eta _{m}>0$: it is indeed
easily seen that these conditions are suf\/f\/icient to guarantee that
$\varphi_{m}\left( x,t\right) $ is \textit{nonsingular} for all
\textit{real} values of the independent variables $x$ and $t$ and
tend asymptotically to the constants $A_{m}$:
\begin{equation*}
\underset{x\rightarrow \pm \infty }{\lim }\left[ \varphi _{m}\left(
x,t\right) \right] =A_{m}.
\end{equation*}
The corresponding solutions $\psi_{n}\left(x,t\right)$ of
the system of \textit{nonlinear} PDEs (\ref{SchrEx1}) are the $N$
zeros of the following polynomial of degree $N$ in the variable
$\psi $:
\begin{gather}
\psi ^{ N}+\sum_{m=1}^{N}\left\{ A_{m}+\frac{B_{m}}{\left(
t+t_{m}+i \eta _{m}\right) ^{ 1/2}} \exp \left[ \frac{-i \left(
x-x_{m}+i \eta _{m} v_{m}\right) ^{ 2}}{4 \left( t+t_{m}+i \eta
_{m}\right) }\right] \right\}  \psi ^{ N-m}  \nonumber \\
\qquad {}=\prod\limits_{n=1}^{N}\left[ \psi -\psi _{n}\left( x,t\right)
\right] . \label{3.3psin}
\end{gather}
 Note that this entails that these solutions satisfy the
asymptotic property (\ref{Asy}), with the constants $a_{n}$ being
the $N$ zeros of the polynomial, of degree $N$ in the variable $a$,
\begin{equation*}
a^{ N}+\sum_{m=1}^{N}A_{m} a^{ N-m}=\prod\limits_{n=1}^{N}\left(
a-a_{n}\right) .
\end{equation*}
Of course the constants $A_{m}$ should be assigned so that
these $N$ zeros $a_{n}$ are \textit{all different}, see~(\ref{Asy}).

A solution of this kind (with $N=3$, and a specif\/ic choice of all
the remaining free parameters) is displayed as an animation in
Section~5.
\end{example}

\subsection{Isochronous version of the class of nonlinear PDEs of Schr\"{o}dinger type}

In this subsection we report, without much commentary, an ``$\omega$-modif\/ied''
 version of (a subclass of) the system of
\textit{nonlinear} PDEs of Schr\"{o}dinger type (\ref{Schr}), which
is characterized by the property to possess an ample class of
solutions \textit{completely periodic} in time with period
\begin{equation}
T=\frac{2 \pi }{\omega }  \label{T}
\end{equation}
(the reason why this is so will be rather obvious from
what follows; for more details on, and other examples of,
\textit{isochronous} PDEs obtained in an analogous manner see
\cite{MC} and Chapter~6 of~\cite{C2006}).

We start from the subcase of the system of \textit{nonlinear} PDEs
of Schr\"{o}dinger type (\ref{Schr}) with
\begin{equation*}
b=0,\qquad W\left( \vec{r}\right) =0,
\end{equation*}
and we set
\begin{equation*}
\tilde{\psi}_{n}\left( \vec{r},t\right) =e^{ i \lambda
 \omega  t}  \psi _{n}\left( \vec{\rho},\tau \right),
\qquad \vec{\rho}\equiv \vec{\rho}\left( t\right) =e^{
\frac{i \omega  t}{2}}  \vec{r},
\qquad \tau \equiv \tau \left(t\right) =\frac{e^{ i \omega  t}
-1}{i \omega },
\end{equation*}
with $\omega$ an arbitrary \textit{positive} constant and
$\lambda$ an \textit{arbitrary} \textit{real rational} number if $a$
vanishes ($a=0$, see (\ref{Schr}) and below), otherwise
$\lambda=-1/2$. It is then easily seen that the new dependent
variables $\tilde{\psi}_{n}\left( \vec{r},t\right)$ satisfy the
following system of \textit{nonlinear} PDEs of Schr\"{o}dinger type:
\begin{equation*}
i \tilde{\psi}_{n,t}-\Delta  \tilde{\psi}_{n}+\lambda  \omega
 \tilde{ \psi}_{n}+\frac{\omega }{2} \vec{r}\cdot
\vec{\nabla} \tilde{\psi} _{n}=2 \sum_{m=1, \,m\neq
n}^{N}\frac{a-\left( \vec{\nabla} \tilde{\psi} _{n}\right) \cdot
\left( \vec{\nabla} \tilde{\psi}_{m}\right) }{\tilde{\psi}
_{n}-\tilde{\psi}_{m}}.  
\end{equation*}
Note that this system is \textit{autonomous} with respect to the time
variable, but it features an \textit{explicit} dependence on the space
variable $\vec{r}$ (see the last term in the left-hand side); and it is
clearly \textit{rotation invariant}.

\begin{example} \label{Example 3.4-1}
The simplest example is again characterized by the assignment
\begin{equation*}
S=1,\qquad a=0,
\end{equation*}
namely by the system of \textit{nonlinear} evolution PDEs
in one-dimensional space (see (\ref{Schr}))
\begin{equation*}
i \tilde{\psi}_{n,t}-\tilde{\psi}_{n,xx}+\lambda  \omega
 \tilde{\psi} _{n}+\frac{\omega
}{2} x \tilde{\psi}_{n,x}=-2 \sum_{m=1,\, m\neq n}^{N}
\frac{\tilde{\psi}_{n,x} \tilde{\psi}_{m,x}}{\tilde{\psi}_{n}-\tilde{\psi}
_{m}}.
\end{equation*}

A special class of solutions of this system of \textit{nonlinear} PDEs
obtains, via the formula
\begin{gather}
\tilde{\psi}_{n}\left( x,t\right) =\exp \left( i \lambda  \omega
 t\right)  \psi _{n}\left( \xi ,\tau \right) ,\nonumber\\
 \xi \equiv \xi
\left( t\right) =\exp \left( \frac{i \omega  t}{2}\right)  x,\qquad
\tau \equiv \tau \left( t\right) =\frac{\exp \left( i \omega
 t\right) -1}{i \omega } ,  \label{4.1Trick}
\end{gather}
 from the solutions $\psi_{n}\left(\xi,\tau\right)$ of
(\ref{Solpsi3.3-1a}), now reading
\begin{equation*}
P_{N}\left( \psi \right) +Q_{N-1}\left( \psi \right)  \exp \left[
-i v \left( \xi -v \tau \right) \right]
=\prod\limits_{n}^{N}\left[ \psi -\psi _{n}\left( \xi ,\tau \right)
\right] ,
\end{equation*}
where $P_{N}\left(\psi\right)$ is an arbitrary
\textit{monic} polynomial of degree $N$ and
$Q_{N-1}\left(\psi\right)$ is an arbitrary polynomial of degree
$N-1$, see~(\ref{P}).


Another class of solutions is provided via (\ref{4.1Trick}) from the
solutions $\psi _{n}$ of (\ref{3.3psin}), now reading
\begin{gather*}
\psi ^{ N}+\sum_{m=1}^{N}\left\{ A_{m}+\frac{B_{m}}{\left( \tau
+\tau _{m}+i \eta _{m}\right) ^{ 1/2}} \exp \left[
\frac{-i \left( \xi -\xi _{m}+i \eta _{m} v_{m}\right)
^{ 2}}{4 \left( \tau +\tau _{m}+i \eta
_{m}\right) }\right] \right\}  \psi ^{ N-m}  \nonumber \\
\qquad{}=\prod\limits_{n=1}^{N}\left[ \psi -\psi _{n}\left( \xi ,\tau
\right) \right] . 
\end{gather*}

\end{example}

\section{Solvable systems of nonlinear PDEs of Klein-Gordon type}

In this section we investigate, f\/irstly by analytic techniques and
subsequently via the explicit display of a few of its solutions, the system
of \textit{nonlinear} evolution equations of Klein--Gordon type~(\ref{KG}).
The f\/irst remarks are analogous to those given in the f\/irst part of the
preceding section and are therefore reported below without much commentary
(their proofs are analogous to those given in the preceding section; we also
use occasionally the same notation, conf\/iding that this will cause no
misunderstandings).

\begin{remark} \label{Remark 4.1}
The ``mean f\/ield'' $\bar{\psi}\left( \vec{r},t\right) $ def\/ined by
(\ref{MeanField}) satisf\/ies now the \textit{linear} Klein--Gordon
equation
\begin{equation*}
\bar{\psi}_{tt}-\Delta \bar{\psi}+M^{ 2} \bar{\psi}=0.
\end{equation*}
\end{remark}

\begin{remark} \label{Remark 4.2}
To a set of \textit{localized }solutions $\psi _{n}\left(
\vec{r},t\right) $ of the system of \textit{nonlinear} evolution
equations of Klein--Gordon type (\ref{KG}) characterized by the
asymptotic conditions (\ref{Asymtpsiphia}) there correspond a set of
\textit{localized} solutions of the system of \textit{linear}
Klein--Gordon PDEs (\ref{EqPhiKG}) characterized by the analogous
asymptotic conditions (\ref{Asymtpsiphib}); and, of course,
viceversa. But such localized solutions cause the same kind of
problem discussed in Remark~\ref{Remark 3.1}. Hence in all
the examples considered below we shall try and avoid this problem,
just as indicated in Remark~\ref{Remark 3.2}.
\end{remark}

\begin{remark} \label{Remark 4.3}
It may be convenient to replace the expression (\ref{Ansb}) of the
monic polynomial $\Psi\left(\psi;\vec{r},t\right)$ in terms of its
coef\/f\/icients $\varphi_{m}\left(\vec{r},t\right)$ by the
representation (\ref{PSIGegena}), again with (\ref{PSIGegenb}) but
now with
\begin{equation}
\gamma =-\frac{M^{ 2}+b \left( 2 N-m\right) }{2 b},  \label{PSIGegenKG}
\end{equation}
 instead of (\ref{PSIGegenc}). Then the requirement that
$\Psi\left(\psi;\vec{r},t\right)$ satisfy the PDE (\ref{EqPSIKG}),
and therefore that its $N$ zeros $\psi_{n}\left(\vec{r},t\right)$
satisfy the system of nonlinear PDEs (\ref{KG}), entails that the
$N$ coef\/f\/icients $\chi_{m}\left(\vec{r},t\right)$ satisfy the
(system of) \textit{decoupled} \textit{linear} Klein--Gordon PDEs
\begin{equation}
\chi _{m,tt}-\Delta \chi _{m}+\left[ M^{ 2}-b (m+3)\right]
 m \chi _{m}=0,\qquad m=1,\dots ,N.  \label{DecoupledKG}
\end{equation}
\end{remark}

Analogous developments to those reported in the preceding Section 3 in the
context of the system of Schr\"{o}dinger-like \textit{nonlinear} evolution
PDEs (\ref{Schr}) can now be elaborated in the present context of the system
of Klein--Gordon-like \textit{nonlinear} evolution PDEs (\ref{KG}). Our
presentation below is more terse than in the preceding section, to avoid
repetitions.

\subsection{Solution of the initial-value problem for the system\\ of
nonlinear Klein--Gordon-like PDEs (\ref{KG})}

The \textit{initial-value} problem consists now in the determination of the
solution $\psi _{n}\left( \vec{r},t\right) $ of the system of
Klein--Gordon-like \textit{nonlinear} PDEs (\ref{KG}) corresponding to
\textit{assigned} initial data $\psi _{m}\left( \vec{r},0\right) $ and $\psi
_{m,t}\left( \vec{r},0\right) ,$ $m=1,\dots ,N.$

The \textit{first} step is to determine the corresponding initial data $%
\varphi _{m}\left( \vec{r},0\right) $ and $\varphi _{m,t}\left( \vec{r}%
,0\right) $ of the system of Klein--Gordon-like \textit{linear} PDEs (\ref%
{EqPhiKG}). This is achieved by f\/irstly solving, as above, for the $N$
functions $\varphi _{m}\left( \vec{r},0\right) $ the system (\ref{phinzero}),
and then by solving for the $N$ functions $\varphi _{m,t}\left( \vec{r}%
,0\right) $ the system
\begin{equation*}
\sum_{m=1}^{N}\varphi _{m,t}\left( \vec{r},0\right)  \psi
^{ N-m}=-\sum_{n=1}^{N}\left\{ \psi _{n,t}\left( \vec{r},0\right)
 \prod\limits_{m=1,\, m\neq n}^{N}\left[ \psi -\psi _{m}\left( \vec{r}
,0\right) \right] \right\} .
\end{equation*}

The \textit{second} step is to obtain the solution $\varphi _{m}\left( \vec{r%
},t\right) $ at time $t$ of the system of evolution PDEs (\ref{EqPhiKG}).
Again, the \textit{linear} character of this system of evolution PDEs
provides the main simplif\/ication.

The \textit{third} step is, as above, to obtain the solution $\psi
_{n}\left( \vec{r},t\right) $ from the, now assumedly known, functions $%
\varphi _{m}\left( \vec{r},t\right) $ via the relation (\ref{phipsi}),
amounting again just to the purely algebraic task of f\/inding the $N$ zeros
of a given monic polynomial of degree $N$.

An alternative procedure of solution can be based on the representation (\ref{PSIGegena})
(with (\ref{PSIGegenb}) and (\ref{PSIGegenKG}) rather than (\ref{PSIGegenc})):
this eases the \textit{second} of the steps outlined above, but
makes a~bit less simple the \textit{first} step. The diligent reader will
easily f\/igure out the relevant details.

\subsection{How to manufacture explicit solutions of the system\\ of nonlinear
Klein--Gordon-like PDEs (\ref{KG})}

As above the strategy -- that underlies all the examples discussed
below -- is to identify an \textit{explicit} solution $\varphi
_{m}\left( \vec{r} ,t\right) $ of the system of \textit{linear}
evolution PDEs (\ref{EqPhiKG}), and then to obtain the corresponding
solution $\psi _{n}\left( \vec{r} ,t\right) $\ of the system of
\textit{nonlinear} evolution PDEs (\ref{KG}) via (\ref{phipsi}),
namely by identifying the $N$ zeros $\psi _{n}\left(
\vec{r},t\right) $\ of the monic polynomial of degree $N$ in $\psi $
with coef\/f\/icients $\varphi _{m}\left( \vec{r},t\right) $, see
(\ref{Ansb}).

An alternative route takes as starting point an \textit{explicit}
solution $\chi_{m}\left(\vec{r},t\right)$ of the
(\textit{decoupled}) system of \textit{linear} evolution PDEs
(\ref{DecoupledKG}), and then obtains the corresponding solution
$\psi_{n}\left( \vec{r},t\right)$\ of the system of
\textit{nonlinear} evolution PDEs (\ref{KG}) by identifying the
quantities $\psi_{n}\left(\vec{r},t\right)$ as the $N$ zeros of the
monic polynomial of degree $N$ in $\psi $ given by the expression
(\ref{PSIGegena}) with~(\ref{PSIGegenb}) and~(\ref{PSIGegenKG}).

\subsection{Examples}

\begin{example} \label{Example 4.3-1}
The simplest example is characterized by the assignment
\begin{equation*}
a=b=0,\qquad M=0,
\end{equation*}
namely by the system of \textit{nonlinear} evolution PDEs (see (\ref{KG}))
\begin{equation}
\psi _{n,tt}-\Delta  \psi _{n}=2 \sum_{m=1,\, m\neq
n}^{N}\frac{\psi _{n,t} \psi _{m,t}-\vec{\triangledown}\psi
_{n}\cdot \vec{\triangledown} \psi _{m}}{\psi _{n}-\psi _{m}}.
\label{KGpsi}
\end{equation}

The corresponding system of \textit{linear} PDEs satisf\/ied by the
coef\/f\/icients $\varphi _{m}\left( \vec{r},t\right) $ of the
polynomial $\Psi \left( \psi ;\vec{r},t\right) $ in the variable
$\psi $ of which the solutions $\psi _{n}\left( \vec{r},t\right) $
are the $N$ zeros, see (\ref{PSIPSI}), reads as follows (see
(\ref{EqPhiKG})):
\begin{equation*}
\varphi _{m,tt}-\Delta  \varphi _{m}=0,\qquad m=1,\dots ,N.
\end{equation*}

The general solution of these \textit{linear} PDEs reads
\begin{equation}
\varphi _{m}\left( \vec{r},t\right) =\sum_{k=1}^{K}f_{mk}\left(
\vec{r}-\vec{ u}_{k} t\right) ,  \label{KGphi}
\end{equation}
with the $K N$ functions $f_{mk}\left( \vec{r}\right) $ arbitrary
and the $K$ constant $S$-vectors $\vec{u}_{k}$ having unit length,
$u_{k}=1,$ but being otherwise arbitrary. Of course these solutions
$\varphi_{m}\left(\vec{r},t\right)$ are \textit{localized} if the
arbitrary functions $f_{mk}\left(\vec{r}\right)$ are themselves
\textit{localized}, but (motivated by Remark~\ref{Remark 4.2}) we shall rather consider solutions that tend asymptotically
to \textit{nonvanishing} asymptotic values; and
$\varphi_{m}\left(\vec{r},t\right)$ has the character of a
\textit{traveling wave} if $K=1$.

The corresponding solutions $\psi_{n}\left(\vec{r},t\right)$ of the
system of \textit{nonlinear} PDEs (\ref{KGpsi}) are the $N$ zeros of
the following polynomial of degree $N$ in $\psi$:
\begin{equation*}
\Psi \left( \psi ;\vec{r},t\right) =\psi ^{ N}+\sum_{m=1}^{N}\psi
^{ N-m} \sum_{k=1}^{K}f_{mk}\left( \vec{r}-\vec{u}_{k} t\right)
=\prod\limits_{n}^{N}\left[ \psi -\psi _{n}\left( \vec{r},t\right)
\right] .  
\end{equation*}

Two solutions of this kind (with $S=2$ respectively $S=1$, $N=3$,
and a specif\/ic choice of the remaining free parameters) are
displayed as animations in Section 5.
\end{example}

\begin{example} \label{Example 4.3-2}
An analogous example -- but reported here for simplicity in the
two-dimensional case ($S=2$) -- is characterized by the analogous
assignment
\begin{equation*}
S=2,\qquad a=b=0,\qquad M=0,
\end{equation*}
 namely by the system of \textit{nonlinear} evolution PDEs
in two-dimensional space (see (\ref{Schr}))
\begin{equation*}
\psi _{n,tt}-\psi _{n,xx}-\psi _{n,yy}=2 \sum_{m=1, \, m\neq n}^{N}\frac{\psi
_{n,t} \psi _{m,t}-\psi _{n,x} \psi _{m,x}-\psi _{n,y} \psi _{m,y}}{\psi
_{n}-\psi _{m}}.
\end{equation*}
 The corresponding system of (decoupled) PDEs satisf\/ied by
the coef\/f\/icients $\varphi_{m}\left(x,y,t\right)$ reads
\begin{equation*}
\varphi _{m,tt}-\varphi _{m,xx}-\varphi _{m,yy}=0.
\end{equation*}

A class of regular solutions of this system of PDEs reads
\begin{equation}
\varphi _{m}\left( x,y,t\right) =J_{0}\left( \sqrt{
(x-x_{m})^{2}+(y-y_{m})^{2}}\right) \left[ A_{m} \cos
(t)+B_{m} \sin (t)\right] +C_{m}, \label{KGphiBessel}
\end{equation}
where $A_{m}$, $B_{m}$ and $C_{m}$ are $3 N$ arbitrary
constants (possibly \textit{complex}), and $J_{0}(r)$ is the
zeroth-order Bessel function of the f\/irst kind. A solution of this
kind (with $N=3,$ and a specif\/ic choice of the remaining free
parameters) is displayed as an animation in Section 5.
\end{example}

\subsection{Isochronous version of the class of nonlinear PDEs of
Klein--Gordon type}

In this subsection we report, with even less commentary than in the
(analogous) Subsection~3.4, an ``$\omega$-modif\/ied'' version of (a
subclass of) the system of \textit{nonlinear} PDEs of Klein--Gordon
type (\ref{KG}), which is again characterized by the property to
possess an ample class of solutions \textit{completely periodic} in
time with period $T,$ see (\ref{T}).

Now we start from the subcase of the system of \textit{nonlinear}
PDEs of Klein--Gordon type (\ref{KG}) with
\begin{equation*}
b=0,\qquad M=0,
\end{equation*}
 and we set
\begin{equation*}
\tilde{\psi}_{n}\left( \vec{r},t\right) =e^{ i \lambda  \omega
 t}  \psi _{n}\left( \vec{\rho},\tau
\right),
\qquad \vec{\rho}\equiv \vec{\rho}\left( t\right) =e^{
i \omega  t}  \vec{r} ,
\qquad \tau \equiv \tau \left( t\right) =\frac{e^{ i \omega  t}
-1}{i \omega },
\end{equation*}
with $\omega$ an arbitrary \textit{positive} constant and
$\lambda$ an \textit{arbitrary real rational} number if $a$ vanishes
($a=0$, see (\ref{KG}) and below), otherwise $\lambda=-1$. It is
then easily seen that the new dependent variables
$\tilde{\psi}_{n}\left(\vec{r},t\right)$ satisfy the following
system of \textit{nonlinear} PDEs of Klein--Gordon type:
\begin{gather*}
\tilde{\psi}_{n,tt}-\Delta  \tilde{\psi}_{n}-2 i \omega  (
\vec{r} \cdot \vec{\nabla})  \tilde{\psi}_{n,t}-i \left(
2 \lambda +1\right)  \omega  \tilde{\psi}_{n,t} \nonumber\\%
\qquad {}-\lambda \left(\lambda +1\right)
 \omega ^{2} \tilde{\psi} _{n}- \left( 2\lambda +1\right)
 \omega ^{ 2} (\vec{r}\cdot \vec{ \nabla})
\tilde{\psi}_{n}-\omega ^{2} (\vec{r}\cdot
\vec{\nabla})^{2} \tilde{\psi}_{n}\nonumber \\
\qquad{}=2 \sum_{m=1,\, m\neq n}^{N}\left\{ \frac{a
-\left(\vec{\nabla} \tilde{\psi}_{n}\right) \cdot
\left(\vec{\nabla} \tilde{\psi}_{m}\right)}{\tilde{\psi}_{n}-\tilde{\psi}_{m}} \right.\nonumber\\
\left.\qquad{}+\frac{\left[ \tilde{\psi} _{n,t}-i\lambda \omega
\tilde{\psi}_{n}-i\omega (\vec{r}\cdot
\vec{\nabla})\tilde{\psi}_{n}\right] \left[ \tilde{\psi
}_{m,t}-i\lambda\omega \tilde{\psi}_{m}-i\omega (\vec{r}
\cdot \vec{\nabla}) \tilde{\psi}_{m}\right]}{\tilde{\psi}_{n}-\tilde{\psi}_{m}}\right\}.
\end{gather*}
Note that this system is \textit{autonomous} with respect
to the time variable, but it features an \textit{explicit}
dependence on the space variable $\vec{r}$ (incidentally, in the
last term in the left-hand side, $(\vec{r}\cdot\vec{\nabla})^{2}\equiv
(\vec{r}\cdot\vec{\nabla}) (\vec{r}\cdot\vec{\nabla}) =
S (\vec{r}\cdot\vec{\nabla})+\sum\limits_{k,j=1}^{S}r_{k} r_{j}\nabla_{k} \nabla_{j}$, where
of course $r_{j}$ denotes the $j$-th component of the $S$-vector
$\vec{r}$); and it is clearly \textit{rotation invariant}.


\section{Animations}

In this section we show a few solutions of the Schr\"{o}dinger and
Klein--Gordon type problems treated in the previous sections, displayed as
animations over time.

Let us begin with a brief description of the methodology -- implemented via
a software written using \textit{Mathematica} -- employed to obtain the
numerical results presented below. After assigning the space dimension $S$,
the coef\/f\/icients $\varphi _{n}\left( \vec{r},t\right) $\ (see (\ref%
{EqPhiSchr}) and (\ref{EqPhiKG})) and the monic polynomial $\Psi \left( \psi
;\vec{r},t\right) $\ of degree $N$ in $\psi $ (see (\ref{Ans})), we compute
the $N$ zeros $\psi _{n}\left( \vec{r},t\right) $\ of this polynomial as
follows: f\/irstly we create in the $(S+1)$-dimensional space of the
independent variables $\vec{r},t$ a lattice; next, we use a root-f\/inding
routine to calculate, at an appropriately chosen point of the lattice, the
(generally \textit{complex}) values of the $N$ zeros $\psi _{n}\left( \vec{r}%
,t\right) $\ of the polynomial $\Psi \left( \psi ;\vec{r},t\right) $, taking
them in a generic order; and then we use an iterative root-f\/inding procedure
to calculate the $N$ zeros of the polynomial $\Psi \left( \psi ;\vec{r}%
,t\right) $ at any new point of the lattice, making use of the zeros
previously calculated at the nearest points of the lattice so as to preserve
the same initial ordering of the zeros.

In the following we consider only examples with $N=3$. The
animations are organized as arrays of synchronized subanimations,
showing above the time evolution of the three $\varphi _{m}\left(
\vec{r},t\right) $ functions, and below the corresponding time
evolution of the three $\psi _{n}\left( \vec{r},t\right) $
solutions. In each frame of the animation (namely, for a f\/ixed value
of the time variable), we display the values of the functions
$\varphi _{m}\left( \vec{r},t\right) $ and $\psi _{n}\left(
\vec{r},t\right) $ (or their \textit{absolute} values $\left\vert
\varphi _{m}\left( \vec{r} ,t\right) \right\vert $ and $\left\vert
\psi _{n}\left( \vec{r},t\right) \right\vert $, when these functions
are \textit{complex}), with respect to the $x$ variable in
two-dimensional plots if $S=1$, or with respect to the $x$ and $y$
variables in three-dimensional plots if $S=2$. In this paper we
restrict attention only to cases with space dimension $S=1$ and
$S=2$.

\subsection{Solutions of (\ref{Schr})}

\begin{figure}[t]
\centering
\includegraphics[width=16cm]{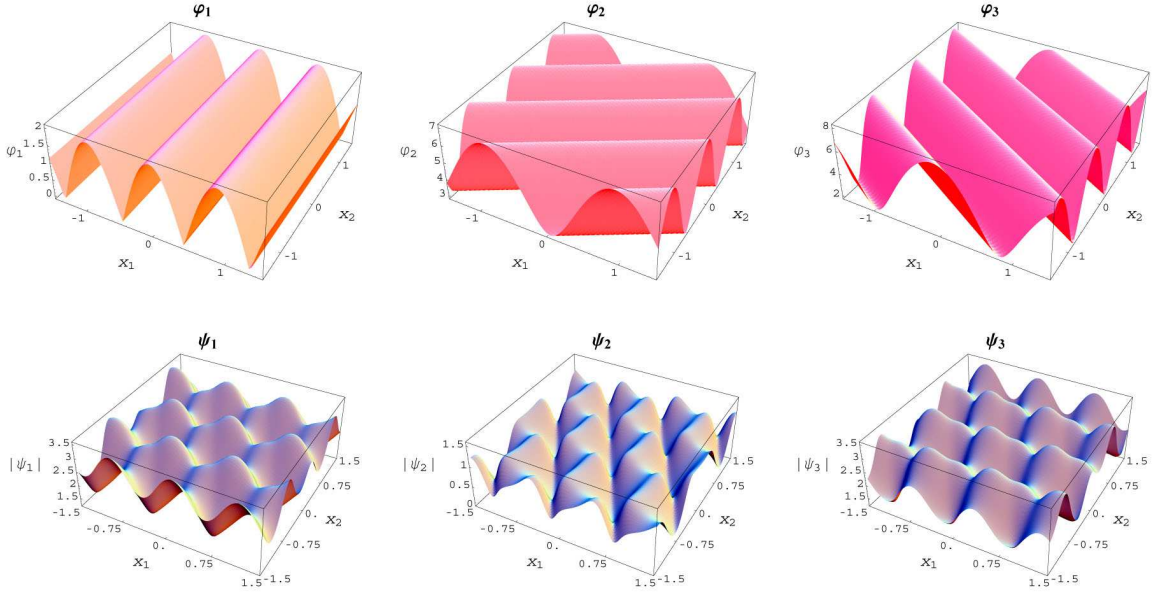}
\caption{The $\protect\varphi _{m}\left( \vec{r},t\right)$ functions
and the absolute values of the $\protect\psi _{m}\left(
\vec{r},t\right)$ solutions for $0\leq t \leq 1/8$.}
\label{Schroedinger_(S=2)_Example1}
\end{figure}

In this subsection we present two numerical solutions of (\ref{Schr})
displayed as animations, the f\/irst one with $S=2$ and the other two with $S=1
$.

\begin{example} \label{Example 5.1-1}
The f\/irst animation 
corresponds
to Example \ref{Example 3.3-1}, with $S=2$, $a=b=0$,
$W(r)=V(r)=0$ and the $\varphi _{m}\left( \vec{r},t\right) $
functions as in (\ref{travelingwave}), namely (but with $3$
dif\/ferent $2$-vector parame\-ters $\vec{v}_{m}$)
\begin{equation*}
\varphi _{m}\left( \vec{r},t\right) =A_{m}+B_{m} \exp \left[ -i \vec{v}
_{m} \left( \vec{r}-\vec{v}_{m} t\right) \right] ,
\end{equation*}
with
\begin{alignat*}{4}
&A_{1}=1 , & & A_{2}=5, & &A_{3}=-5, &  \nonumber \\ %
&B_{1}=1, & &B_{2}=2, & &B_{3}=3, &  \nonumber \\ %
&\vec{v}_{1}=(4 \sqrt{\pi }, 0), \qquad & &\vec{v}_{2}=(-2  \sqrt{\pi }, 2 \sqrt{3 \pi}) , \qquad
& &\vec{v}_{3}=(-2 \sqrt{\pi}, -2 \sqrt{3 \pi }).&
\end{alignat*}
Here the periods in time of the three
$\varphi_{m}\left(\vec{r},t\right)$ functions are chosen to be the
same, $T=1/8$, and the animation is performed on the time interval
$0\leq t\leq T$, then closed in loop.

Fig.~\ref{Schroedinger_(S=2)_Example1} is the f\/irst frame.
To see the whole animation, please click on the following (external) link:
\href{http://www.emis.de/journals/SIGMA/2006/Paper088/Animation5.1.1.gif}{http://www.emis.de/journals/SIGMA/2006/Paper088/Animation5.1.1.gif}.
\end{example}

\begin{figure}[t]
\centering
\includegraphics[width=16cm]{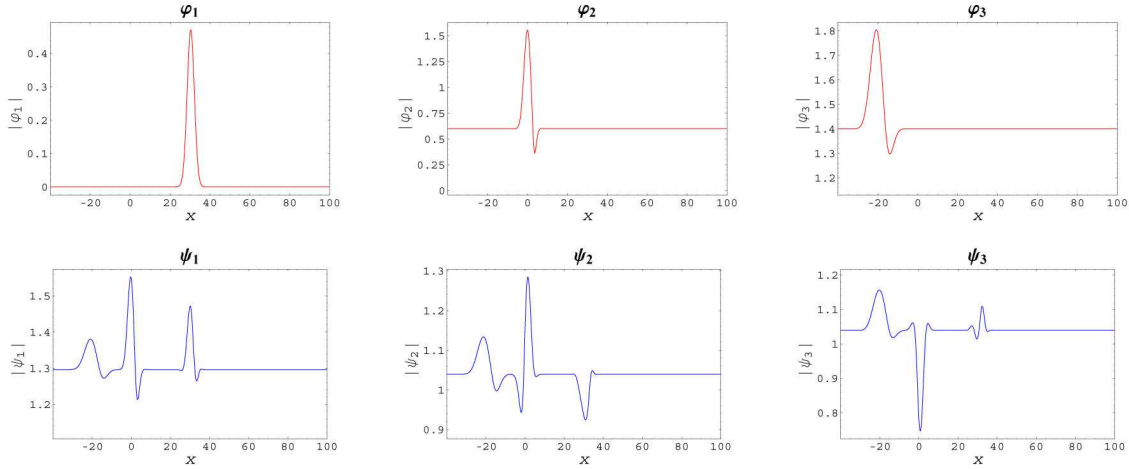}
\caption{The absolute values of the $\protect\varphi _{m}\left(
x,t\right)$ functions and the absolute values of the
$\protect\psi_{m}\left( x,t\right)$ solutions for $0\leq t \leq
20$.} \label{Schroedinger_(S=1)_Example2}
\end{figure}

\begin{example} \label{Example 5.1-2}
The second animation 
corresponds
to the Example \ref{Example 3.3-1}, with $S=1$, $a=b=0$,
$W(r)=V(r)=0$ and the $\varphi_{m}\left(x,t\right)$ functions as in
(\ref{localizedphi}), namely
\begin{equation*}
\varphi _{m}\left( x,t\right) =A_{m}+\frac{B_{m}}{\left( t+t_{m}+i \eta
_{m}\right) ^{ 1/2}} \exp \left[ \frac{-i \left( x-x_{m}+i \eta
_{m} v_{m}\right) ^{ 2}}{4 \left( t+t_{m}+i \eta _{m}\right) }\right],
\end{equation*}
 with
\begin{alignat*}{7}
&A_{1}=0, & &B_{1}=0.5,& &x_{1}=30, & &v_{1}=0.5, & &\eta _{1}=1, & &t_{1}=0.8, &  \nonumber \\
&A_{2}=-0.6,\qquad & &B_{2}=-0.9,\qquad & &x_{2}=0,& &v_{2}=1,& &\eta _{2}=2,& &t_{2}=0.4,&  \nonumber \\
&A_{3}=1.4,& &B_{3}=0.8,& &x_{3}=-20,\qquad & &v_{3}=0.5,\qquad & &\eta _{3}=7, \qquad & &t_{3}=1.&
\end{alignat*}
 Here the animation is performed on the time interval
$0\leq t\leq 20$.

Fig.~\ref{Schroedinger_(S=1)_Example2} is the f\/irst frame.
To see the whole animation, please click on the following (external) link:
\href{http://www.emis.de/journals/SIGMA/2006/Paper088/Animation5.1.2.gif}{http://www.emis.de/journals/SIGMA/2006/Paper088/Animation5.1.2.gif}.
\end{example}

\subsection{Solutions of (\ref{KG})}

In this subsection we present three numerical solutions of (\ref{KG})
displayed as animations, two with $S=2$ and the last with $S=1$.

\begin{figure}[t]
\centering
\includegraphics[width=16cm]{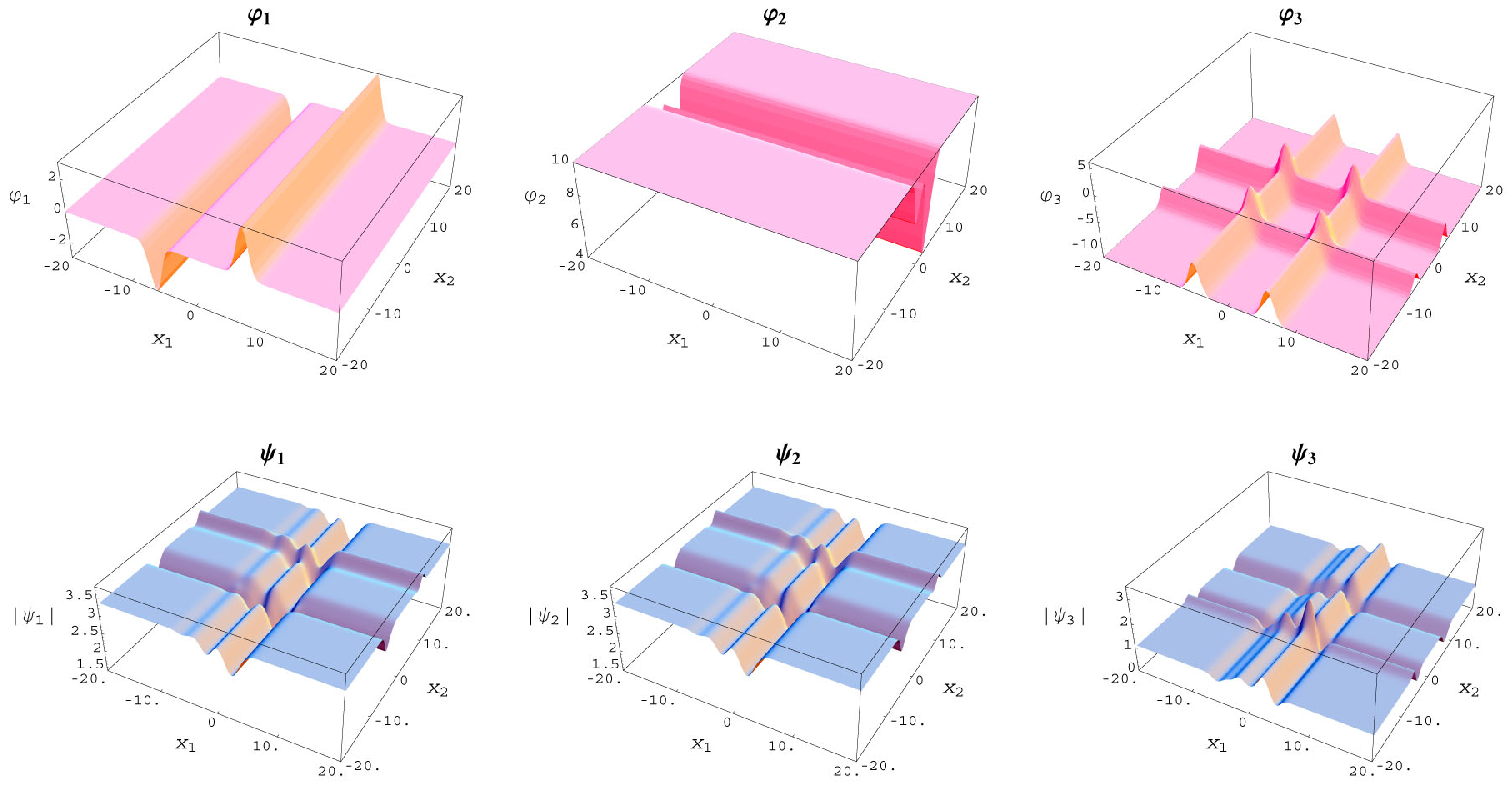}
\caption{The $\protect\varphi _{m}\left( \vec{r},t\right)$ functions
and the absolute values of the $\protect\psi_{m}\left(
\vec{r},t\right)$ solutions for $0\leq t \leq 20$.}
\label{Klein-Gordon_(S=2)_Example1}
\end{figure}

\begin{example} \label{Example 5.2-1}
The f\/irst animation 
corresponds
to the Example \ref{Example 4.3-1}, with $S=2$, $a=b=0$,
$M=0$ and the $\varphi_{m}\left(x,t\right)$ functions as in
(\ref{KGphi}), with $K=4$ and a very particular choice of the
functions $\varphi_{m}\left(x,y,t\right)$:
\begin{gather*}
\varphi _{m}\left( x,y,t\right)=A_{m} \exp \left[ -\frac{\left(
x-x_{m}^{(1)}-t\right) ^{ 2}}{a_{m}}\right] +B_{m} \exp \left[
-\frac{
\left( x-x_{m}^{(2)}+t\right) ^{ 2}}{b_{m}}\right] \nonumber \\
\phantom{\varphi _{m}\left( x,y,t\right)=}{}+C_{m} \exp \left[ -\frac{\left( y-y_{m}^{(1)}-t\right)
^{ 2}}{c_{m}} \right] +D_{m} \exp \left[ -\frac{\left(
y-y_{m}^{(2)}+t\right) ^{ 2}}{ d_{m}}\right]+E_{m}  ,
\end{gather*}
where
\begin{alignat*}{6}
&A_{1}=-3, & &B_{1}=3, & &C_{1}=0, & &D_{1}=0, & &E_{1}=0, &  \notag \\%
&A_{2}=0, & &B_{2}=0, & &C_{2}=-6, & &D_{2}=-3, & &E_{2}=10, &  \notag \\%
&A_{3}=6, & &B_{3}=6, & &C_{3}=3, & &D_{3}=4.5, & &E_{3}=-12, &  \notag \\%
&a_{1}=1.5, & &b_{1}=1.3, & &c_{1}=1.3, & &d_{1}=1.6, & && \notag \\%
&a_{2}=1.8, & &b_{2}=1.2, & &c_{2}=2, & &d_{2}=1.6, & && \notag \\%
&a_{3}=1.4, & &b_{3}=1.5, & &c_{3}=1.4, & &d_{3}=1.2, & && \notag \\%
&x_{1}^{(1)}=-6, & &x_{1}^{(2)}=7, & &y_{1}^{(1)}=0, & &y_{1}^{(2)}=0, & && \notag \\%
&x_{2}^{(1)}=0, & &x_{2}^{(2)}=0, & &y_{2}^{(1)}=2.5, & &y_{2}^{(2)}=-1.5,\qquad  & && \notag \\%
&x_{3}^{(1)}=-5,\qquad  & &x_{3}^{(2)}=6, \qquad & &y_{3}^{(1)}=-5.5,\qquad  & &y_{3}^{(2)}=4.5. & && %
\end{alignat*}
Here the animation is performed on the time interval
$0\leq t\leq 20$.

Fig.~\ref{Klein-Gordon_(S=2)_Example1} is the f\/irst frame.
To see the whole animation, please click on the following (external) link:
\href{http://www.emis.de/journals/SIGMA/2006/Paper088/Animation5.2.1.gif}{http://www.emis.de/journals/SIGMA/2006/Paper088/Animation5.2.1.gif}.
\end{example}

\begin{figure}[t]
\centering
\includegraphics[width=16cm]{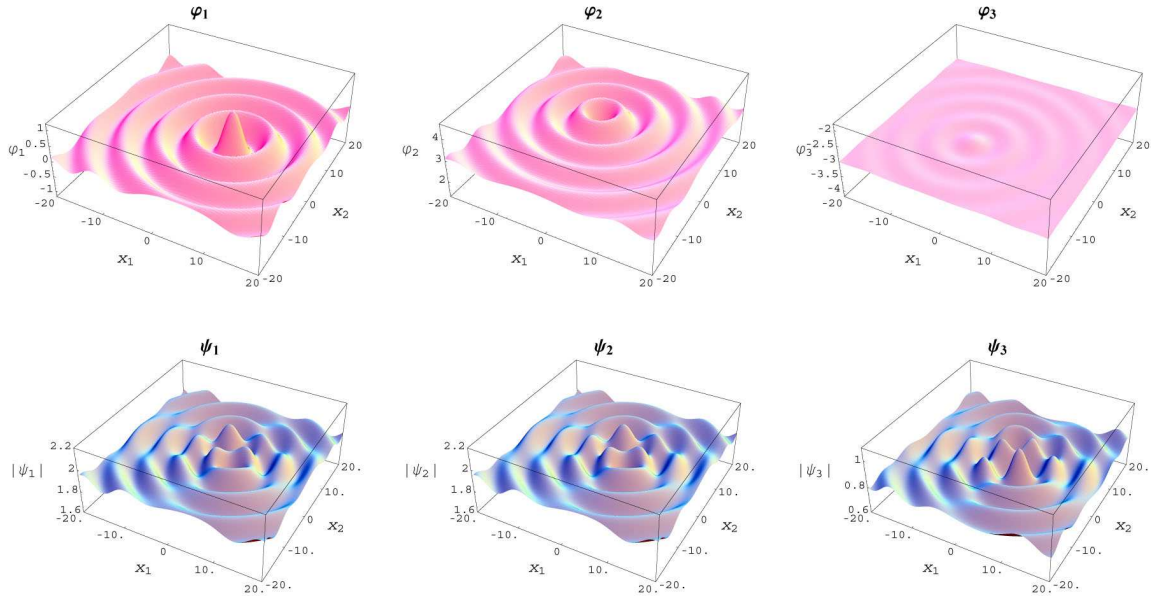}
\caption{The $\protect\varphi _{m}\left( \vec{r},t\right) $
functions and the absolute values of the $\protect\psi _{m}\left(
\vec{r},t\right) $ solutions for $0\leq t\leq 2 \protect\pi $.}
\label{Klein-Gordon_(S=2)_Example2}
\end{figure}

\begin{example} \label{Example 5.2-2}
The second animation 
corresponds to the Example~\ref{Example 4.3-2}, with $S=2$,
$a=b=0$, $M=0$ and the $\varphi _{m}\left( \vec{r},t\right) $
functions as in (\ref{KGphiBessel}):
\begin{equation*}
\varphi _{m}\left( x,y,t\right) =J_{0}\left( \sqrt{
(x-x_{m})^{2}+(y-y_{m})^{2}}\right) \left[ A_{m} \cos
(t)+B_{m} \sin (t)\right] +C_{m}, 
\end{equation*}
where
\begin{alignat*}{6}
&A_{1}=1, & &B_{1}=0.1,\qquad & &C_{1}=0, & &x_{1}=5, & &y_{1}=0, &  \notag \\%
&A_{2}=-1, & &B_{2}=1, & &C_{2}=3, & &x_{2}=-5/2, & &y_{2}=5 \sqrt{3}/2, &  \notag \\%
&A_{3}=0.1,\qquad  & &B_{3}=1, & &C_{3}=-3,\qquad & &x_{3}=-5/2, \qquad & &y_{3}=-5 \sqrt{3}/2. &  %
\end{alignat*}
The $\varphi_{m}\left(\vec{r},t\right)$ functions are
periodic in time with period $T=2 \pi$. The animation is performed
on the time interval $0\leq t\leq T$, then closed in loop.

Fig.~\ref{Klein-Gordon_(S=2)_Example2} is the f\/irst frame.
To see the whole animation, please click on the following (external) link:
\href{http://www.emis.de/journals/SIGMA/2006/Paper088/Animation5.2.2.gif}{http://www.emis.de/journals/SIGMA/2006/Paper088/Animation5.2.2.gif}.
\end{example}

\begin{figure}[t]
\centering
\includegraphics[width=16cm]{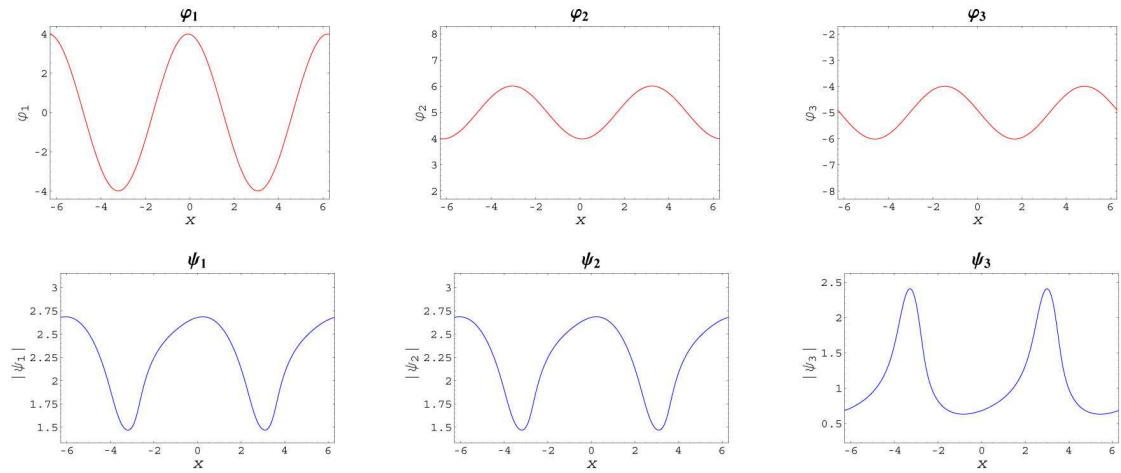}
\caption{The $\protect\varphi _{m}\left( x,t\right) $ functions and
the absolute values of the $\protect\psi _{m}\left( x,t\right) $
solutions for $ 0\leq t\leq 2 \protect\pi $.}
\label{Klein-Gordon_(S=1)_Example3}
\end{figure}

\begin{example} \label{Example 5.2-3}
The third animation 
corresponds
again to the Example~\ref{Example 4.3-1}, with $S=1$, \mbox{$a=b=0$}, $M=0$ and
the $\varphi_{m}\left(x,t\right)$ functions as in (\ref{KGphi}) with
$K=2$, and with a very particular choice of the functions
$\varphi_{m}\left(x,t\right)$:
\begin{equation*}
\varphi _{m}\left( x,t\right) =A_{m} \cos (x-t+B_{m})+C_{m} \cos
(x+t+D_{m})+E_{m},
\end{equation*}
where
\begin{alignat*}{6}
&A_{1}=1, & &B_{1}=0, & &C_{1}=3, & &D_{1}=0.1, & &E_{1}=0, &  \notag \\%
&A_{2}=2, & &B_{2}=\pi, & &C_{2}=-1,\qquad & &D_{2}=\pi +0.1, \qquad & &E_{2}=5, &  \notag \\%
&A_{3}=-1,\qquad & &B_{3}=\pi/2+0.1,\qquad & &C_{3}=2, & &D_{3}=\pi/2, & &E_{3}=-5. &  %
\end{alignat*}
The $\varphi_{m}\left(x,t\right)$ functions
are periodic in time with period $T=2 \pi$ and periodic with
respect to the $x$ variable with period $L=\pi$. The animation is
performed on the time interval $0\leq t\leq T$, then closed in loop.

Fig.~\ref{Klein-Gordon_(S=1)_Example3} is the f\/irst frame.
To see the whole animation, please click on the following (external) link:
\href{http://www.emis.de/journals/SIGMA/2006/Paper088/Animation5.2.3.gif}{http://www.emis.de/journals/SIGMA/2006/Paper088/Animation5.2.3.gif}.
\end{example}

\section{Outlook}

In future articles we plan to report additional investigations of
these solvable (systems of) nonlinear evolution PDEs, and in
particular to display visual animations of solutions of certain of
these models in \textit{three-dimensional} space.

\subsection*{Acknowledgements}
The final correction of this paper was done
while both authors were taking part in a Scientific Gathering on Integrable
Systems and the Transition to Chaos at the Centro International de Ciencias in
Cuernavaca.

\LastPageEnding

\end{document}